\documentclass[conference]{IEEEtran}
\IEEEoverridecommandlockouts
\usepackage{amsmath,amssymb,amsfonts}
\usepackage{algorithmic}
\usepackage{algorithm}
\usepackage{array}
\usepackage{textcomp}
\usepackage{stfloats}
\usepackage{url}
\usepackage{verbatim}
\usepackage{graphicx}
\usepackage{cite}
\usepackage[subtle,tracking=normal]{savetrees}
\usepackage{subcaption}
\usepackage{xcolor}
\usepackage{caption}
\usepackage{ulem}
\usepackage[left=0.673in,right=0.667in,top=0.75in,bottom=1.1in]{geometry}
\usepackage{makecell}

\begin{document}
\title{Impact~of~Objective~Function~on~Spectral~Efficiency in Integrated HAPS-Terrestrial Networks}
\author{\IEEEauthorblockN{Afsoon Alidadi Shamsabadi\IEEEauthorrefmark{1}, Animesh Yadav\IEEEauthorrefmark{2}, and Halim Yanikomeroglu\IEEEauthorrefmark{1}}
\IEEEauthorblockA{\IEEEauthorrefmark{1}Carleton University,  Ottawa, ON, Canada, and \IEEEauthorrefmark{2}Ohio University, Athens, OH, USA\\
Email: afsoonalidadishamsa@sce.carleton.ca, yadava@ohio.edu, halim@sce.carleton.ca}}
\maketitle
\begin{abstract}
Integrating non-terrestrial networks (NTNs), in particular high altitude platform stations (HAPS), with terrestrial networks, referred to as vHetNets, emerges as a promising future wireless network architecture for providing ubiquitous connectivity. In this context, optimizing the performance of vHetNets has become a paramount concern, particularly in harmonized spectrum vHetNets, where HAPS and terrestrial networks share the same frequency band, resulting in severe inter-/intra-tier interference. This paper provides a comparative analysis of different objective functions, specifically focusing on weighted sum rate (WSR), network-wide proportional fairness (NW-PF), and network-wide max-min fairness (NW-MMF), with an aim to design a joint user association scheme and multiple-input multiple-output (MIMO) beamforming weights in a vHetNet, operating in an urban area. The simulation results comprehensively compare the behavior of different objective functions in vHetNets and standalone terrestrial networks. This analysis aims to shed light on the impact of diverse objective functions on the achievable spectral efficiency (SE) of vHetNets.
\end{abstract}
\vspace{-2mm}
\section{Introduction}
In the context of next-generation wireless communications, the transition to sixth-generation (6G) networks presents a critical challenge in achieving ubiquitous and seamless connectivity. This transition demands innovative architectures to satisfy growing coverage and heterogeneous data traffic. To this end, one promising approach is integrating aerial platforms, particularly high altitude platform stations (HAPS), located approximately $20$ kilometers above the ground, with terrestrial networks to form vertical heterogeneous network (vHetNet) architecture for future telecommunications networks\cite{HAPSSurvey,VHetNet,HAPS_Mag}. There are several advantages and challenges that differentiate vHetNets from terrestrial heterogeneous networks (HetNets). Advantages include the large footprint of HAPS, and strong line-of-sight (LoS) connection between HAPS and user equipment (UEs); challenges include limited wireless backhaul capacity and onboard energy caused by the positioning of HAPS. Moreover, in harmonized spectrum vHetNets, where HAPS-tier and terrestrial network-tier share the same frequency band, inter-/intra-tier interference is a critical challenge, which necessitates developing efficient algorithms to mitigate and manage propagated interference.
The algorithm design must ensure to enhance the network's overall performance by optimizing the key network parameters such as radio resources, antenna array geometry, or other related aspects of wireless networks. Diverse optimization problems can be formulated in this context, each with a distinct objective and impact on network performance. The network objective functions range from improving the performance of the worst UE to enhancing the average or total achievable data rate or spectral efficiency (SE); choosing the most appropriate objective function is important.

The widely used objective functions to study the performance of wireless networks include weighted sum rate (WSR), network-wide proportional fairness (NW-PF), and network-wide max-min fairness (NW-MMF). Particularly, WSR maximizes the weighted sum of all UEs' data rate, NW-PF maximizes the data rate of individual UE, providing a trade-off among achieved data rate and fairness\cite{PF}, and NW-MMF maximizes the minimum achievable data rate of all UEs across the entire network\cite{PF}. It is worth mentioning that the term network-wide (NW) is used to emphasize that NW-PF and NW-MMF objective functions consider the entire network for achieving fairness; although, other variations consider achieving per-cell fairness. Several works have designed interference management schemes in vHetNets, utilizing various objective functions \cite{HAPSIM_SR,UA,myletter_WCL,myletter_CL}. The authors in \cite{HAPSIM_SR} proposed an interference coordination method in integrated HAPS-terrestrial networks to maximize the sum data rate of the UEs. In \cite{UA}, the authors designed a user association scheme based on the deep Q-learning (DQL) approach to maximize the network's sum data rate. In \cite{myletter_WCL}, we employed the NW-MMF objective function to design joint subcarrier and power allocation in a vHetNet. Furthermore, in \cite{myletter_CL}, we solved the NW-MMF problem to design a joint user association scheme and beamforming weights in a vHetNet. 

In standalone terrestrial networks, because of non-LoS (NLoS) links and interference, some UEs face severely low data rate. In this case, fairness-based objective functions would improve the performance of impacted UEs. On the other hand, in standalone HAPS networks, with dominant LoS link, WSR would be an appropriate objective function. Whereas for the vHetNets, it is unclear that which objective function will be appropriate. Hence, this work investigates the impact of objective function on SE performance of vHetNets. To this end, this study formulates three joint user association and beamforming weights design (JUBD) problems corresponding to three objective functions, namely WSR, NW-PF, and NW-MMF, in a vHetNet. Since the formulated problems are strictly non-convex, reformulations and approximations are employed to transform the original problems into tractable forms, that can be solved iteratively through successive convex approximation (SCA) framework\cite{SCA}. Comprehensive comparisons are provided through numerical simulations to analyze the impact of the three objective functions on SE in a vHetNet.

The remainder of the paper is organized as follows. Section \ref{System Model} presents the system model, Section \ref{Sec:Algorithm} formulates the optimization problems, and explains the proposed algorithms, and Section \ref{Sec:Convergence} discusses the convergence behavior and complexity of the proposed algorithm. Section \ref{Sec:Results} presents the numerical results obtained, and finally, Section \ref{Sec:Conclusion} concludes the paper.
\vspace{-1mm}
\section{System Model}\label{System Model}
\begin{figure}[t]
    \centering
    \captionsetup{justification=centering}
    \includegraphics[width=0.65\linewidth]{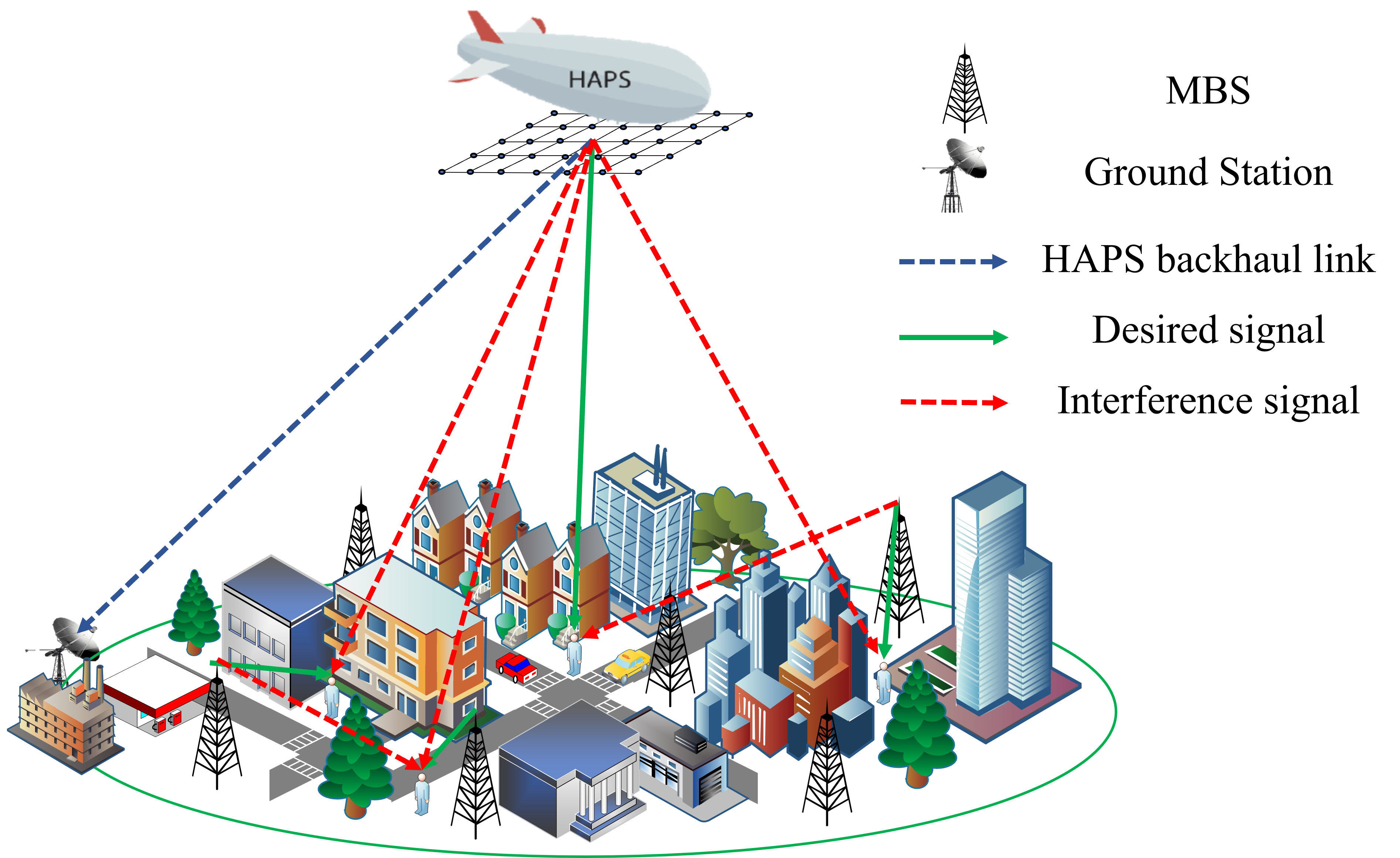}
    \caption{\small Network architecture.}
    \label{fig_1}
\end{figure}

Consider a vHetNet consisting of one HAPS and $B$ terrestrial macro base stations (MBSs), providing service in a downlink channel to $U$ single antenna UEs within an urban geographical region, as illustrated in Fig.~\ref{fig_1}. The base stations\footnote{In this work, the term base station refers to HAPS and MBSs, collectively.} and UEs are indexed by $b\in\{1,\ldots,B+1\}$ and $u\in \{1,\ldots,U\}$, respectively, with index $b=B+1$ reserved for HAPS. Each base station $b$ is equipped with a total of $N_b$ antenna elements that are arranged in a uniform planar array (UPA) geometry. Base stations employ beamforming to generate narrow high gain beams toward UEs; the beamforming matrix at base station $b$ is represented by $\mathbf{W}^{b}=[\mathbf{w}^b_1,\dots,\mathbf{w}^b_U] \in \mathbb{C}^{N_b \times U}$, where $\mathbf{w}^b_u=[w^b_{1,u},\dots,w^b_{N_b,u}]^T \in \mathbb{C}^{N_b}$, is defined as the beamforming weight vector of UE $u$ at base station $b$. Furthermore, we define $\mathbf{H}^b=[\mathbf{h}^b_1,\dots,\mathbf{h}^b_U] \in \mathbb{C}^{N_b \times U}$, as the channel matrix at base station b, where $\mathbf{h}^b_u=[h^b_{1,u},\dots,h^b_{N_b,u}]^T \in \mathbb{C}^{N_b}$ denotes the channel vector between base station $b$ and UE $u$. In this system model, the same time-frequency resource, with the same bandwidth, has been allocated to all UEs, resulting in propagated interference from all active base stations toward all UEs.

Considering the different locations of HAPS and MBSs, different propagation characteristics are expected within HAPS and terrestrial tiers. Particularly, due to the altitude of HAPS, there is a high probability of having a strong LoS connection between HAPS and UEs, aligned with the small-scale fading components, inherited from the multi-path nature of urban environments. In this regard, $\mathbf{h}^{B+1}_u$, representing the channel vector between HAPS and UE $u$, can be modeled as a 3D Rician fading channel with a dominant LoS and NLoS components as \cite{HAPS-MIMO}
\vspace{-2mm}
\begin{equation}\label{HAPS-Channel}
\small
    \mathbf{h}_u^{B+1}=\frac{1}{\sqrt{{PL}_{(B+1),u}}}\Big(\sqrt{\frac{1}{1+K_u}}\hat{\mathbf{h}}_u+\sqrt{\frac{K_u}{1+K_u}}\overline{\mathbf{h}}_u\Big),~\forall u,\qquad \qquad
\end{equation}
where ${PL}_{(B+1),u}$ represents the free~space~path~loss~(FSPL) between HAPS and UE $u$, calculated as ${PL}_{b,u}=(4\pi f_c d_{b,u}/{c})^2,~\forall u,~b=B+1,$ where $f_c$ is the carrier frequency (in Hz), $d_{b,u}$ (in m) is the distance between the base station $b$ and UE $u$, and $c$ represents the speed of light in free space. Atmospheric loss is negligible due to the operating frequency of the vHetNet (sub-6 GHz). $K_u$ is the Rician factor for UE $u$, and $\hat{\mathbf{h}}_u \in \mathbb{C}^{N_{B+1}}$ represents the NLoS component of the channel vector with its elements from a normal random distribution with zero mean and unit variance, $\mathcal{NC}(0,1)$. Accordingly, $\overline{\mathbf{h}}_u$ is the LoS component of the channel vector obtained from the steering vectors\cite{HAPS-MIMO}.

In contrast, for the MBS-UE channel in an urban environment, there is a very low probability of having LoS connection; therefore, we consider small-scale fading, FSPL, and shadowing for the channel between each MBS and UE. In this regard, the channel coefficient between antenna element $r$ of MBS $b$ and UE $u$, denoted as $h^{b}_{r,u}$, can be formulated as $h^b_{r,u}={\hat{h}^b_{r,u}\xi^b_u}/{\sqrt{{PL}_{b,u}}}$, where $\hat{h}_{r,u}^b\sim \mathcal{NC}(0,1)$, represents the small-scale Rayleigh fading channel gain, and $\xi^b_u=10^{{{\xi_u^{'b}}}/10}$ denotes the log-normal shadowing gain, where ${\xi_u^{'b}}$ is the Gaussian random variable with zero mean and standard deviation $\sigma_\xi$ (in dB).
In this system, we represent the binary user association matrix by $\mathbf{A}\in~\mathbb{B}^{(B+1)\times U}$, with elements $a_{b,u}=1,~\forall b,~\forall u,$ if UE $u$ is associated to base station $b$, and $a_{b,u}=0,$ otherwise. Accordingly, the signal-to-interference-plus-noise ratio (SINR) for UE $u$ can be defined as
\begin{equation}\label{eq:SINR}
    \gamma_u=\cfrac{\sum_{b=1}^{B+1}{\left|\left(\mathbf{h}^{b}_{u}\right)^H \mathbf{w}^{b}_{u}\right|^2}}{\sum_{b=1}^{B+1}\sum_{\substack{k=1 \\ k\not=u}}^{U}{\left|\left(\mathbf{h}^{b}_{u}\right)^H \mathbf{w}^{b}_{k}\right|}^2+\sigma^2_n},~\forall u,
\end{equation}
where $\sigma^2_n$ denotes the variance of the additive white Gaussian noise (AWGN) at UE $u$, and $(\cdot)^H$ refers to the conjugate transpose operation. It is noteworthy that the equation \eqref{eq:SINR} is applicable under the assumption that the beamforming weights for non-associated UEs are zero at each base station.
\vspace{-1mm}
\section{Problem Formulation and Proposed Algorithm}\label{Sec:Algorithm}
In this section, we formulate three JUBD problems for three different objective functions: WSR, NW-PF, and NW-MMF. Due to the non-convexity of the original problems, we apply reformulations and approximations to convert them into tractable and approximate forms. These approximated SCA-based problems can be solved iteratively for the optimal solution, which is essentially the suboptimal solution of the corresponding original problem.
\subsection{WSR JUBD Algorithm}
The WSR JUBD problem aims to maximize the weighted sum of the UEs' data rate. Considering the assumption that all the UEs are allocated the same amount of bandwidth, we continue the discussion by replacing the data rate of each UE with SE. Defining $0 \leq \xi_u \leq 1$ as the weight, associated with UE $u$, the preliminary WSR JUBD optimization problem can be expressed as
\vspace{-2mm}
\begin{IEEEeqnarray*}{lcl}\label{eq:WSR}
    &\underset{\mathbf{A},\mathbf{W}}{\text{maximize}}\,\, ~& \sum_{u=1}^{U} {\xi_u\log_2(1+\gamma_u)} \,  \IEEEyesnumber \IEEEyessubnumber* \label{eq:WSR_Obj}\\
    &\text{s.t.} & \gamma_u \geq \gamma_{\text{min}},~\forall u, \qquad \label{eq:WSR_const0}\\
    && \sum_{u=1}^U a_{(B+1),u} F \log_2(1+\gamma_u) \leq {R_{\text{BH}}}, \qquad \label{eq:WSR_const1}\\
    && \|\mathbf{w}^b_{u}\|^2_2 \leq P^{\text{max}}_b a_{b,u},\,~\forall u,~\forall b, \qquad \label{eq:WSR_const2}\\
    && {\|\mathbf{W}^b \|^2_F}\leq P^{\text{max}}_b, \, ~ \forall b,\qquad \label{eq:WSR_const3}\\
    && \sum_{b=1}^{B+1}{a_{b,u}} = 1, \,  ~\forall u, \label{eq:WSR_const4}
\end{IEEEeqnarray*}
where $\gamma_u$ represents the SINR of UE $u$, calculated according to \eqref{eq:SINR}, $\mathbf{W}$ denotes the collection of all beamforming matrices $\mathbf{W}^b,~\forall b$, and $\gamma_{\text{min}}$ refers to the minimum required SINR for each UE. $F$ denotes the bandwidth, allocated to each UE, $R_{\text{BH}}$ refers to the available data rate of the HAPS backhaul link, and $P_b^{\text{max}}$ is the maximum available transmit power at base station $b$. Problem \eqref{eq:WSR} is a mixed integer non-linear programming (MINLP) maximization problem under constraints \eqref{eq:WSR_const0}-\eqref{eq:WSR_const4}. Constraint \eqref{eq:WSR_const0} ensures that the minimum SINR requirement of each UE $u$ is satisfied, and constraint \eqref{eq:WSR_const1} limits the sum of the data rate of associated UEs with HAPS to $R_\text{BH}$. Constraint \eqref{eq:WSR_const2} ensures that the beamforming weights for non-associated UEs at each base station are zero, constraint \eqref{eq:WSR_const3} limits the total transmit power of each base station $b$ to $P_b^{\text{max}}$, and finally, constraint \eqref{eq:WSR_const4} ensures that each UE is associated with only one base station. Due to non-convex objective function, non-convex constraints \eqref{eq:WSR_const0} and \eqref{eq:WSR_const1}, and the existence of binary variables $\mathbf{A}$, problem \eqref{eq:WSR} is challenging to be solved for the optimal solution. Consequently, through the steps discussed in the following, we transform problem \eqref{eq:WSR} into an approximated SCA-based convex form, which can be solved iteratively. An on-the-shelf algorithm, such as the interior point method, can be used to find the optimal solution at each SCA iteration. Note that the solution to the approximated problem is essentially a suboptimal solution of the original problem \eqref{eq:WSR}.

\textbf{Step 1:} First, we deal with the objective function. To tackle this, considering the monotonically increasing behavior of the logarithm function, we replace the objective function by $\prod_{u=1}^{U}{(1+\gamma_u)^{\xi_u}}$. Next, we define slack variables $\mathbf{t}=[1,\dots,t_U]^T,$ with the elements $t_u,~\forall u,$ as the lower bound for ${(1+\gamma_u)}^{\xi_u}$, and replace the objective function by its lower bound, $f(\mathbf{t})=\prod_{u=1}^{U}{t_u}$. Accordingly, $U$ constraints will be added to the problem as 
\begin{equation}\label{WSR_New_Const_1}
    {t_u}^{1/\xi_u} \leq {1+\gamma_u},~\forall u.
\end{equation}
\begin{figure*}[t]
    \small 
    \begin{IEEEeqnarray*}{lcl}\label{WSR_SOCP constraints}
        \|[2z^{(Q)}_i~~t_{(2i-1)}-t_{(2i)}]^T\|_2 \leq t_{(2i-1)}+t_{(2i)},~i=\{1,\dots,2^{Q-1}\},\IEEEyesnumber \IEEEyessubnumber* \qquad\\
        \|[2z^{(q)}_i~~z^{(q+1)}_{(2i-1)}-z^{(q+1)}_{(2i)}]^T\|_2 \leq z^{(q+1)}_{(2i-1)}+z^{(q+1)}_{(2i)},~q=\{1,\dots,Q-1\},~i=\{1,\dots,2^{q-1}\}.   
    \end{IEEEeqnarray*}
    \hrule
\end{figure*}
The objective function $f(\mathbf{t})$ can be reformulated as a second-order cone programming (SOCP) representation\cite{SOCP}. In this regard, we consider the special case of $U=2^Q$, where $Q$ is a positive integer\footnote{If $U \neq 2^Q$, we can consider nearest power of $2$ that is greater than $U$, and define $t_u=1,~\text{for}~u \in \{U+1,\dots,2^Q\}$.}. Accordingly, we define ${2^Q}-1$ new slack variables $z^{(q)}_i,~\forall q \in \{1,\dots,Q\},~i \in \{1,\dots,2^{q-1}\}$. Consequently, the objective function can be replaced by $z^{(1)}_1$, and new series of SOC constraints will be added to the problem according to \eqref{WSR_SOCP constraints}, which is mentioned at the top of this page. 

\textbf{Step 2:} In this step, we tackle the constraint \eqref{WSR_New_Const_1} which is non-convex due to the fractional formula of $\gamma_u$. To handle this, we define slack variables $\beta_u,~\forall u$, and $\alpha_u,~\forall u$, and replace $\gamma_u$ in constraint \eqref{WSR_New_Const_1} by its lower bound $\alpha_u$, as
\begin{equation}\label{eq:WSR_New_Const_1_V2}
    {t_u}^{1/{\xi_u}} \leq 1+\alpha_u,~\forall u,
\end{equation}
that represents a convex constraint for $0 \leq \xi_u \leq 1$.
In order to ensure that $\alpha_u$ is the lower bound for $\gamma_u$, two new constraints should be defined as
\begin{IEEEeqnarray*}{lcl}\label{betaalpha}
    \alpha_u \beta_u \leq \sum_{b=1}^{B+1}{\left|\left(\mathbf{h}^{b}_{u}\right)^H \mathbf{w}^{b}_{u}\right|}^2,~\forall u, \IEEEyesnumber \IEEEyessubnumber* \label{eq:alphabetaileqPRi}\\
    \beta_u \geq \sum_{b=1}^{B+1}\sum_{\substack{k=1 \\ k\not=u}}^{U}{\left|\left(\mathbf{h}^{b}_{u}\right)^H \mathbf{w}^{b}_{k}\right|}^2+\sigma^2_n,\,~\forall u. \label{eq:betaigeqIi}
\end{IEEEeqnarray*}

Constraint \eqref{eq:betaigeqIi} is convex but constraint \eqref{eq:alphabetaileqPRi} is non-convex, due to the norm function in the right-hand side. To handle this, we define slack variable matrices $\mathbf{P} \in \mathbb{R}^{(B+1) \times U},$ and $\mathbf{Q} \in \mathbb{R}^{(B+1) \times U},$ with their elements, respectively, defined as
\begin{IEEEeqnarray*}{lcl}\label{Real and Imag}
    p_{b,u} \leq \Re{\left(\left(\mathbf{h}^{b}_{u}\right)^H \mathbf{w}^{b}_{u}\right)},~\forall u,~\forall b, \qquad \IEEEyesnumber \IEEEyessubnumber* \label{Real}\\
    q_{b,u} \leq \Im{\left(\left(\mathbf{h}^{b}_{u}\right)^H \mathbf{w}^{b}_{u}\right)},~\forall u,~\forall b, \qquad\label{Imag}
\end{IEEEeqnarray*}
where, $\Re (\cdot)$ and $\Im (\cdot)$ represent the real and imaginary parts of a complex number, respectively.
Consequently, equation \eqref{eq:alphabetaileqPRi} can be reformulated as 
\begin{equation}\label{eq:alphaleqV2}
    \alpha_u \beta_u \leq {\sum_{b=1}^{B+1}{(p_{b,u})^2+(q_{b,u})^2}},~\forall u,
\end{equation}
which can be converted to a convex form, according to \eqref{first Taylor}, by moving $\beta_u$ to the right-hand side, and replacing the right-hand side with its approximated first-order Taylor series:
\begin{multline}{\label{first Taylor}}
\small
     \alpha_u \leq \sum_{b=1}^{B+1}\frac{2{p^{(n)}_{b,u}}}{{\beta^{(n)}_{u}}}\left(p_{b,u}-{p^{(n)}_{b,u}}\right)+\frac{2{q^{(n)}_{b,u}}}{{\beta^{(n)}_u}}\left(q_{b,u}-{q^{(n)}_{b,u}}\right)\\ +\frac{\left({p^{(n)}_{b,u}}\right)^2+\left({q^{(n)}_{b,u}}\right)^2}{\beta^{(n)}_u}\left(1-\frac{\beta_u-{\beta^{(n)}_u}}{{\beta^{(n)}_u}}\right),~\forall u.
\end{multline}

In equation \eqref{first Taylor}, the superscript $(n)$ refers to the value of the corresponding variable in the $n$th iteration of the SCA process. 

\textcolor{black}{\textbf{Step 3:} Through this step, we deal with constraint \eqref{eq:WSR_const1}. To this end, first, we prove that at each SCA iteration $n$, SINR for all UEs, i.e, $\gamma_u,~\forall u,$ will converge to $\alpha_u,~\forall u$, upon convergence.}

\textcolor{black}{\textit{Proof:} We assume that $t^{(n)}_u$ and $\alpha^{(n)}_u,$ are the optimal solutions at SCA iteration $n$. By contradiction, we show that at the convergence of SCA iteration $n$, the term $(t_u^{(n)})^{1/\xi_u}-1$ converges to $\alpha^{(n)}_u$; in other words, constraint \eqref{eq:WSR_New_Const_1_V2} will be active. If constraint \eqref{eq:WSR_New_Const_1_V2} is not active upon convergence, i.e, $(t_u^{(n)})^{1/\xi_u}-1 < \alpha^{(n)}_u$, there will be $\hat{t}_u > t^{(n)}_u,~\forall u$, which can be considered as the better solution than $t^{(n)}_u$. As this contracts with the assumption, we conclude that at the convergence, the constraint \eqref{eq:WSR_New_Const_1_V2} will be active. Now, according to inequalities $(t_u^{1/\xi_u})-1 \leq \gamma_u \leq \alpha_u,~\forall u,$ and the squeeze theorem, $\gamma_u,~\forall u,$ also converges to $\alpha^{(n)}_u,~\forall u$.} 

\textcolor{black}{Consequently, by replacing $\gamma_u$ with $\alpha_u$, employing Jensen's inequality\footnote{$\mathbb{E}\{\log(f(x))\} \leq \log(\mathbb{E}\{f(x)\})$, where $\mathbb{E}$ denotes the expected value.}, and defining slack variables $\eta_u=a_{(B+1),u} \alpha_u,~\forall u$, we can replace constraint \eqref{eq:WSR_const1} as}
\begin{equation}\label{eq:BH_2}
    \sum_{u=1}^U \eta_u \leq {2^{R_{\text{BH}}/{F}}-1},
\end{equation}
which represents a convex constraint. Accordingly, four new constraints will be added to the problem, as below, to ensure that the actual equality $\eta_u=a_{(B+1),u} \alpha_u,~\forall u$ holds:
\begin{IEEEeqnarray*}{lcl}\label{eq:BH_1}
    0\leq \eta_u \leq \gamma_{\text{max}} a_{(B+1),u},~\forall u,\IEEEyesnumber \IEEEyessubnumber* \label{new_a}\\
    0\leq \alpha_u-\eta_u \leq \gamma_{\text{max}}(1-a_{(B+1),u}),~\forall u,\label{new_b}
\end{IEEEeqnarray*}
where $\gamma_{\text{max}}$ is large number denoting the upper bound for $\alpha_u$.
As a result, the approximate problem of problem \eqref{eq:WSR} will be as
\begin{IEEEeqnarray}{lcl}\label{eq:WSR_V2}
    &\underset{\mathbf{A},\mathbf{W},\mathbf{t},\mathbf{z},\boldsymbol{\beta},\boldsymbol{\alpha},\mathbf{P},\mathbf{Q},\boldsymbol{\eta}}{\text{maximize}}\,\, & z_1^{(1)} \,  \IEEEyesnumber \IEEEyessubnumber* \label{eq:P2_Obj}\\
    &\text{s.t.} & \eqref{eq:WSR_const2},\eqref{eq:WSR_const3},\eqref{eq:WSR_const4},\eqref{WSR_SOCP constraints},\eqref{eq:WSR_New_Const_1_V2},\\
    && \eqref{eq:betaigeqIi},\eqref{Real and Imag},\eqref{first Taylor},\eqref{eq:BH_2},\eqref{eq:BH_1},\qquad
\end{IEEEeqnarray}
where $\mathbf{z},~\boldsymbol{\beta},~\boldsymbol{\alpha},~\text{and}~\boldsymbol{\eta}$ are the vectors, which collect variables $ z_i^{(q)},~\beta_u,~\alpha_u,~\text{and}~\eta_u,$ respectively. 

\subsection{NW-PF JUBD Algorithm}
Adopting the logarithm utility function for the NW-PF objective function, the preliminary formulated NW-PF JUBD problem can be expressed as
\begin{IEEEeqnarray*}{lcl}\label{eq:PF}
    &\underset{\mathbf{A},\mathbf{W}}{\text{maximize}}\,\, & ~\sum_{u=1}^{U} {\log(\log(1+\gamma_u))} \,  \IEEEyesnumber \IEEEyessubnumber* \label{eq:PF_Obj}\\
    &\text{s.t.} & \eqref{eq:WSR_const1},\eqref{eq:WSR_const2},\eqref{eq:WSR_const3},\eqref{eq:WSR_const4}.
\end{IEEEeqnarray*}

Similar to problem \eqref{eq:WSR}, problem \eqref{eq:PF} is a MINLP due to the presence of binary variable $\mathbf{A}$, non-convex objective function and non-convex constraint \eqref{eq:WSR_const1}. Therefore, we convert the problem to the approximate SCA-based tractable form, to be solved iteratively. To deal with the objective function, we convert the summation to the product as $\prod_{u=1}^{U} \log(1+\gamma_u)$. Accordingly, $U$ slack variables $\mathbf{t}=[t_1,\dots,t_U]^T$ will be defined as the lower bound for $\log_2(1+\gamma_u),~\forall u$. In this way, the objective function can be replaced by
$f(\mathbf{t})=\prod_{u=1}^{U}{t_u}$,
that can be reformulated as a SOC presentation, according to Step 1 in problem \ref{eq:WSR} relaxation. As a result, the set of SOC constraints \eqref{WSR_SOCP constraints} will be added to the problem, and the objective function will be replaced by $z_1^{(1)}$. Moreover, $U$ new constraints will be added to the problem as 
\begin{equation}\label{exp_const}
    t_u \leq \log(1+\gamma_u),~\forall u.
\end{equation}

Similar to the relaxation approach, proposed in Step 2 of problem \eqref{eq:WSR} relaxation, we replace $\gamma_u$ with its lower bound $\alpha_u$, and add constraints \eqref{eq:betaigeqIi}, \eqref{Real and Imag}, and \eqref{first Taylor} to problem \eqref{eq:PF}. In this way, constraint \eqref{exp_const} will be replaced by
\begin{equation}{\label{eq:PF_Const1_new}}
     e^{t_u} \leq 1+\alpha_u,~\forall u,
\end{equation}
which can be represented as a series of SOC constraints as \eqref{Exp-cone}, utilizing the exponential cone approximation\cite{Exp-cone}. In \eqref{Exp-cone}, parameter $m$ determines the accuracy of the approximation, and $k^u_i,~i=1,\dots,m+4,~\forall u,$ are the newly defined slack variables. 
\begin{IEEEeqnarray*}{lcl}\label{Exp-cone}
    k^u_{m+4} \leq 1+\alpha_u,~\forall u,\IEEEyesnumber \IEEEyessubnumber* \label{Exp-cone1}\\
    \|[2+{t_u/2^{(m-1)}}~~1-k^u_1]^T\|_2 \leq 1+k^u_1,~\forall u,\label{Exp-cone2}\\
    \|[5/3+{t_u/2^{m}}~~1-k^u_2]^T\|_2 \leq 1+k^u_2,~\forall u,\label{Exp-cone3}\\
    \|[2k^u_1~~1-k^u_3]^T\|_2 \leq 1+k^u_3,~\forall u,\label{Exp-cone4}\\
    19/72+k^u_2+1/24k^u_3 \leq k^u_4,~\forall u,\label{Exp-cone5}\\
    \|[2k^u_{i-1}~~1-k^u_i]^T\|_2 \leq 1+k^u_i,~i=5,6,\dots,m+4,~\forall u.\qquad\label{Exp-cone6}
\end{IEEEeqnarray*}

Finally, we use the technique, presented in Step 3 of problem \eqref{eq:WSR} relaxation, to transform constraint \eqref{eq:WSR_const1} to an equivalent convex form.
Consequently, problem \eqref{eq:PF} can be approximated as
\begin{IEEEeqnarray}{lcl}\label{eq:PF_V2}
    &\underset{\mathbf{A},\mathbf{W},\mathbf{k},\mathbf{z},\boldsymbol{\beta},\boldsymbol{\alpha},\mathbf{P},\mathbf{Q},\boldsymbol{\eta}}{\text{maximize}}\,\, & z_1^{(1)} \,  \IEEEyesnumber \IEEEyessubnumber* \label{eq:PF_Obj_V2}\\
    &\text{s.t.} & \eqref{eq:WSR_const2},\eqref{eq:WSR_const3},\eqref{eq:WSR_const4},\eqref{WSR_SOCP constraints},\eqref{eq:betaigeqIi},\qquad\\
    && \eqref{Real and Imag},\eqref{first Taylor},\eqref{eq:BH_2},\eqref{eq:BH_1},\eqref{Exp-cone},\qquad
\end{IEEEeqnarray}
where $\mathbf{k}$ is the vector of slack variables $k^u_i,~\forall u,~i=1,\dots,m+4$.

\subsection{NW-MMF JUBD Algorithm}
The NW-MMF JUBD problem, maximizes the minimum SE of the UEs, which is equivalent to maximizing the minimum SINR of the UEs within the entire network. In this way, the NW-MMF JUBD problem can be formulated as
\begin{IEEEeqnarray*}{lcl}\label{eq:NW-MMF}
    &\underset{\mathbf{A},\mathbf{W}}{\text{maximize}}\,\, & ~\underset{u}{\min ~\gamma_u} \,  \IEEEyesnumber \IEEEyessubnumber* \label{eq:NWMMF_Obj}\\
    &\text{s.t.} & \eqref{eq:WSR_const1},\eqref{eq:WSR_const2},\eqref{eq:WSR_const3},\eqref{eq:WSR_const4}.
\end{IEEEeqnarray*}

Problem \eqref{eq:NW-MMF} can be reformulated by replacing the objective function with new variable ${\gamma}_m$, representing the minimum SINR of the entire network, and adding $U$ constraints according to \eqref{eq:MMF_const1} to the problem to ensure that the MMF is maintained within the entire network. 
\begin{equation}\label{eq:MMF_const1}
    \gamma_u \geq \gamma_m,~\forall u.
\end{equation}

With a similar rationale as the WSR and NW-PF problems \eqref{eq:WSR} and \eqref{eq:PF}, problem \eqref{eq:NW-MMF} represents a MINLP that can be converted to equivalent SCA form as \eqref{eq:NW-MMF_2} using the techniques introduced in steps 2 and 3 of the WSR problem:
\begin{IEEEeqnarray*}{lcl}\label{eq:NW-MMF_2}
    &\underset{\mathbf{A},~\mathbf{W},\gamma_m,\boldsymbol{\beta},\boldsymbol{\alpha},\mathbf{P},\mathbf{Q},\boldsymbol{\eta}}{\text{maximize}}\,\, & ~\gamma_m \,  \IEEEyesnumber \IEEEyessubnumber* \label{eq:NWMMF_Obj2}\\
    &\text{s.t.} & \alpha_u \geq \gamma_m, \forall u,\\
    && \eqref{eq:WSR_const2},\eqref{eq:WSR_const3},\eqref{eq:WSR_const4},\eqref{eq:betaigeqIi},\eqref{Real and Imag},\eqref{first Taylor},\eqref{eq:BH_2},\eqref{eq:BH_1}. \qquad
\end{IEEEeqnarray*}

Algorithm~\ref{alg:JUBD} describes the JUBD algorithm for three discussed objective functions. Algorithm~\ref{alg:JUBD} solves the approximated problem \eqref{eq:WSR_V2}, \eqref{eq:PF_V2}, or \eqref{eq:NW-MMF_2}, iteratively for the optimal solution. Upon convergence, which is defined as when a tolerable change in the objective function value is achieved (less than $\epsilon$), or the maximum number of SCA iterations, $N_\text{iter}$, is reached, whichever first, the algorithm stops.
\begin{algorithm}[h!]
\caption{\small Iterative WSR/NW-PF/NW-MMF JUBD algorithm for vHetNet.}\label{alg:JUBD}
    \begin{algorithmic}[1]
    \small
    \STATE \textbf{Input:}~$U,~B,~N_{b},~\mathbf{H}^b,~P_b^{\text{max}},~\xi_u,~\gamma_\text{min},~\gamma_\text{max},~F,~R_\text{BH},~\sigma^2_n,~N_{\text{iter}}$.\\
    \STATE \textbf{Output:} $\mathbf{W}^*,~\mathbf{A}^*$.\\
    \STATE \text{Initialize} $\boldsymbol{\beta}^{(0)},~\mathbf{P}^{(0)},~\mathbf{Q}^{(0)},$ \text{and set} $n:=0.$\\
    \STATE \textbf{while} {not Converged} \textbf{do}\\
    \STATE \hspace{0.3cm} \text{Solve \eqref{eq:WSR_V2}/\eqref{eq:PF_V2}/\eqref{eq:NW-MMF_2} to find $\mathbf{A}^{(n)*},~\mathbf{W}^{(n)*},~\boldsymbol{\beta}^{(n)*},~\mathbf{Q}^{(n)*},$}\\
    \hspace{0.3cm} \text{$\mathbf{P}^{(n)*},~\boldsymbol{\gamma}_m^{(n)*}$}.\\
    \STATE \hspace{0.3cm} \text{Update $\boldsymbol{\beta}^{(n+1)}:=\boldsymbol{\beta}^{(n)*},~\mathbf{P}^{(n+1)}:=\mathbf{P}^{(n)*},\mathbf{Q}^{(n+1)}:=\mathbf{Q}^{(n)*},$}\\
    \hspace{0.4cm}\text{$n:=n+1$}.
    \end{algorithmic}
\label{alg_JUBD}
\end{algorithm}
\begin{table}[!t]
\caption{\small Simulation Parameters.}\label{tab:table1}
\centering

\begin{tabular}{|c||c|}
\hline
\textbf{Parameter} & \textbf{Value}\\
\hline
\makecell{Shadowing standard deviation, $\sigma_\xi$,\\ Rician factor,~$K_u,~\forall u$} & $8,~10$\\
\hline
Noise variance, $\sigma^2_n$ & $-100$ dBm\\
\hline
\makecell{Number of antenna elements, $N_{\text{b}},~b\in \{1,\dots,B\}$,\\~$N_{B+1}$} & $4 \times 4$,~$8 \times 8$\\
\hline
\makecell{Available transmit power, $P_b^{\text{max}},~b\in \{1,\dots,B\}$,\\~$P_{B+1}^{\text{max}}$} & $43$ dBm,~$52$ dBm\\
\hline
UE allocated bandwidth, $F$ & $1$ MHz \\
\hline
HAPS backhaul link available data rate,~$R_\text{BH}$ & $20$ Gbps \\
\hline
UE minimum SINR requirement,~$\gamma_\text{min}$ & $1000$\\
\hline
Algorithm~\ref{alg:JUBD} convergence parameters, $N_{\text{iter}},~\epsilon$ & $10,~10^{-4}$\\
\hline
\end{tabular}
\end{table}
\vspace{-2mm}
\section{Convergence and Complexity Analysis}\label{Sec:Convergence}
In this section, we discuss the convergence and complexity of NW-MMF JUBD algorithm. The discussion can be generalized to other two algorithms with the same rationale. First, it is evident that $\gamma^{(n)}_m$ is a feasible point for approximated problems at iteration $(n+1)$. In accordance with the optimality condition, $\gamma^{(n)}_m \leq \gamma^{(n+1)}_m$, it confirms the monotonically increasing behavior of the SCA problem's objective function across iterations. Further, the approximated problems are inherently bounded maximization problems due to the imposition of the maximum transmit power constraint. Thus, the sequence of iterate $\gamma_m^{(n)}$ is convergent. Finally, the optimal point of the SCA-based approximated problem satisfies the Karush-Kuhn-Tucker (KKT) conditions of the original problem \cite[Proposition 3.2]{SCA}. The worst-case computational cost of the algorithm is mainly determined by the branch-and-bound method to solve mixed-integer problem at each SCA iteration and is given as $\mathcal{O}(2^{U\left(B+1\right)})$.
\section{Numerical Results}\label{Sec:Results}
This study considers a $4$ Km by $4$ Km square urban geographical area, covered by one HAPS and four MBSs, serving $16$ uniformly distributed UEs, under the carrier frequency $f_c=2.545$ GHz. 
Simulation results compare the statistical behavior of per UE SE (i.e., $\log_2(1+\gamma_u),~\forall u$), sum SE (i.e., $\sum_{u=1}^{U}\log_2(1+\gamma_u)$), and minimum SE (i.e., $\underset{u}\min~\gamma_u$). These performance metrics are analyzed under the implementation of the JUBD Algorithm 1, tailored for WSR, NW-PF, and NW-MMF objectives in the envisioned vHetNet. In the WSR scenario, equal weights are assumed for all UEs, i.e., $\xi_u=1,~\forall u$, resulting in sum rate (SR) scenario. In addition to vHetNet, we consider a standalone terrestrial network, comprising $5$ MBSs; accordingly, we implement JUBD Algorithm~\ref{alg:JUBD} in this standalone terrestrial network and compare the results with the vHetNet. The rest of the simulation parameters are listed in Table~\ref{tab:table1}. For each scenario, outcomes are derived from $1000$ independent and identically distributed (i.i.d.) random placements of the 16 UEs within the area.

\begin{figure*}[t]
    \centering
    \captionsetup{justification=centering}
    \begin{subfigure}{0.65\columnwidth}
        \centering
        \includegraphics[width=\columnwidth]{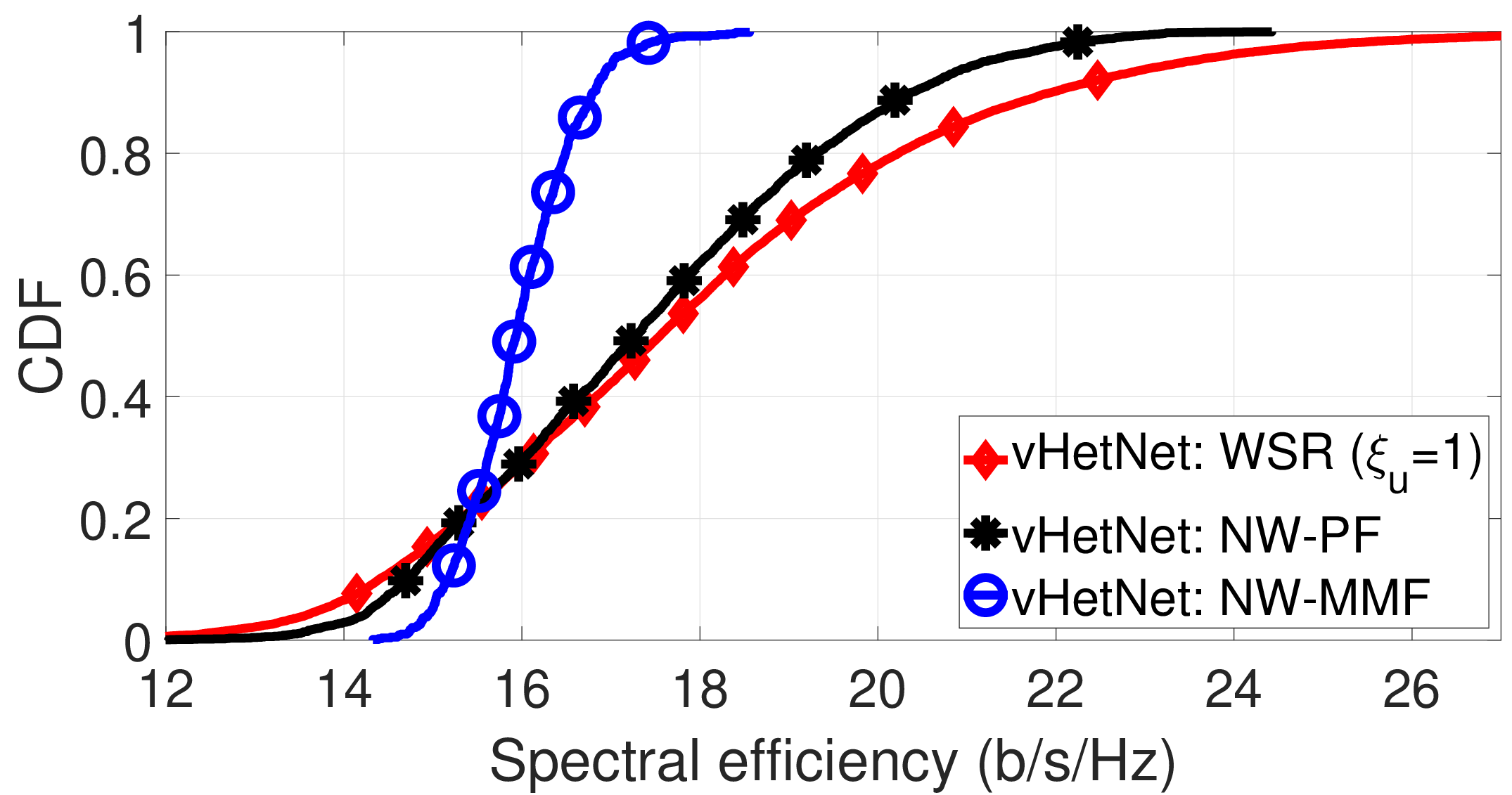}
        \caption{\small CDF of per UE SE.}
        \label{fig_2:a}
    \end{subfigure}
    \begin{subfigure}{0.65\columnwidth}
        \centering
        \includegraphics[width=\columnwidth]{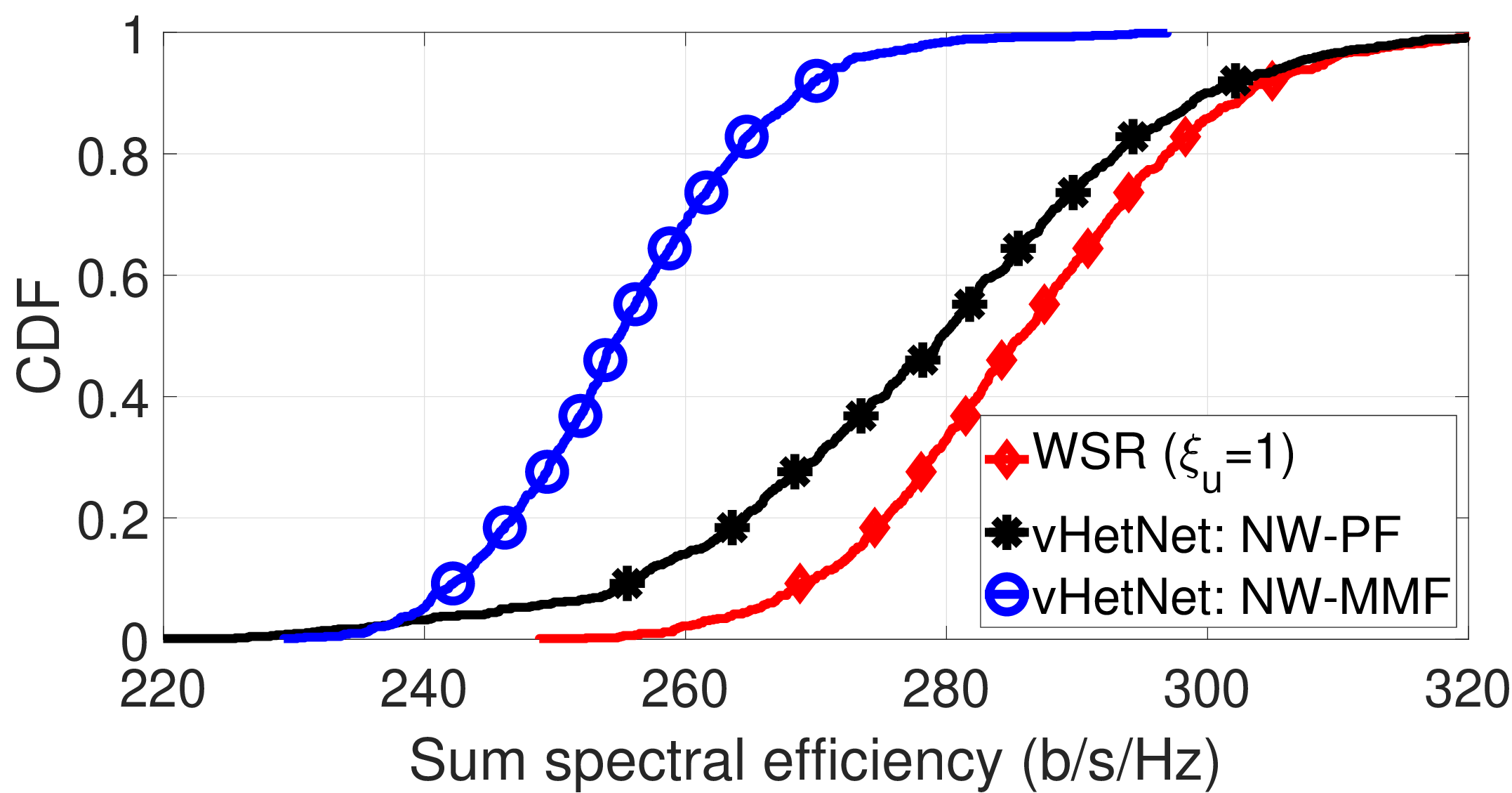}
        \caption{\small CDF of sum SE.}
        \label{fig_2:b}
    \end{subfigure}
    \begin{subfigure}{0.65\columnwidth}
        \centering
        \includegraphics[width=\columnwidth]{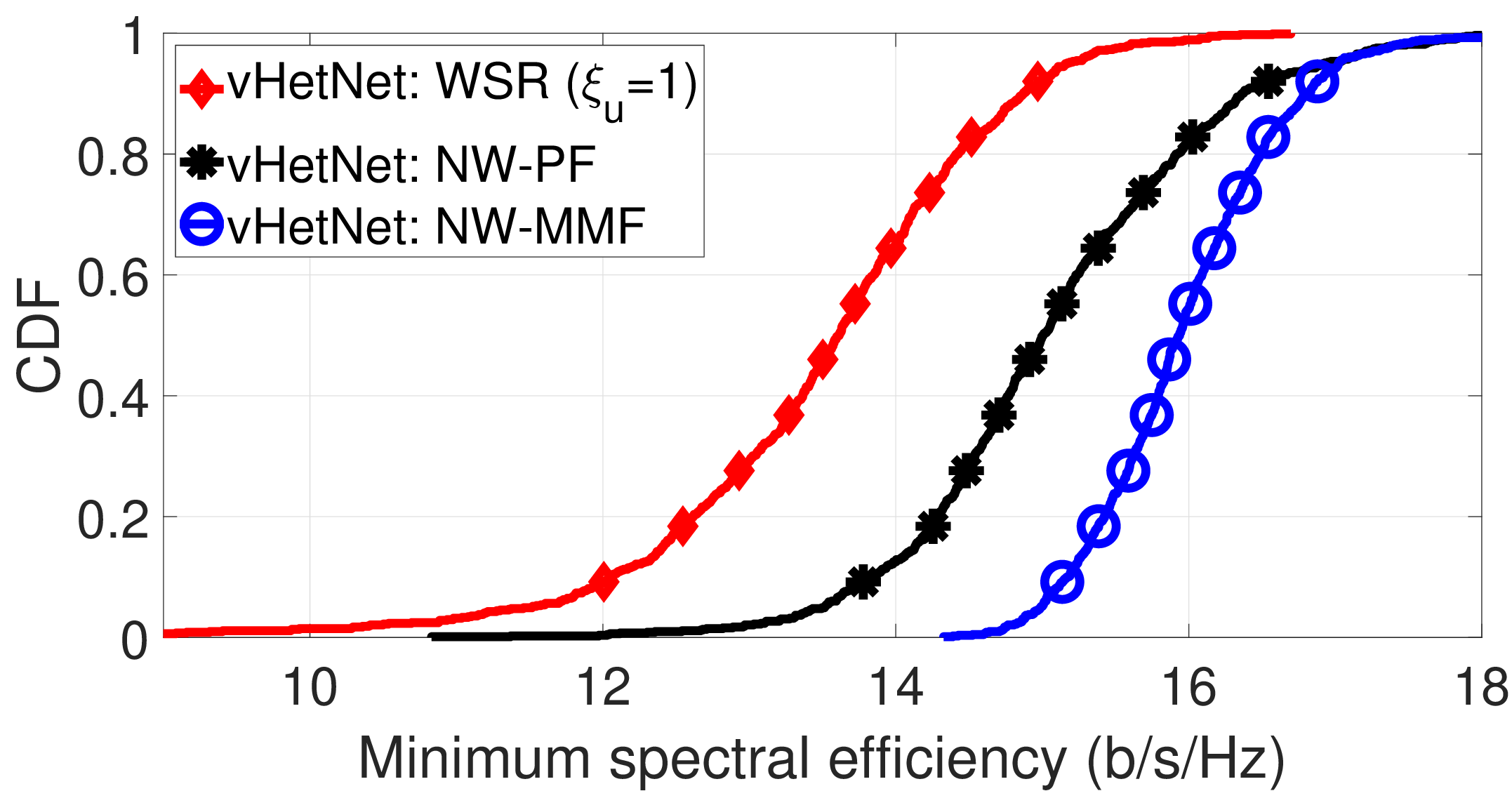}
        \caption{\small CDF of minimum SE.}
        \label{fig_2:c}
    \end{subfigure}
    \vspace{-1.5mm}
    \caption{\small Statistical behavior of SE in a vHetNet (4 MBSs + 1 HAPS) for different objective functions.}
    \label{fig_2}
\end{figure*}

Fig.~\ref{fig_2} illustrates the cumulative distribution function (CDF) of per UE SE, sum SE, and minimum SE, obtained by implementing JUBD Algorithm~\ref{alg:JUBD} in the vHetNet. Two observations can be deduced from Fig.~\ref{fig_2:a}, which plots the CDF of per UE SE. First, it can be observed that NW-PF and WSR objective functions lead to higher average SE compared to NW-MMF; on the other hand, NW-MMF objective function achieves a higher fifth percentile value compared to NW-PF and WSR scenarios, providing an improved performance for the worst UE. Second, the CDF curve in NW-MMF has a steeper curve compared to the other two objective functions, validating a lower variance for per UE SE values in NW-MMF scenario. The second observation validates the more stable quality of service (QoS) for UEs in different locations across the region.

According to Fig.~\ref{fig_2:b}, WSR algorithm provides the highest sum SE. In addition, according to Fig.~\ref{fig_2:c}, the NW-MMF objective function provides the highest minimum SE. Furthermore, NW-PF can be considered as a trade-off between sum SE and minimum SE. Specifically, NW-PF provides a higher sum SE compared to NW-MMF and provides a higher minimum SE compared to WSR scenario. Therefore, while choosing proper objective function for JUBD Algorithm~\ref{alg:JUBD} in vHetNet depends on the specific requirement of the network (improving worst UE performance, achieve higher sum data rate, or providing fairness), it can be concluded from the results in Fig.~\ref{fig_2} that using NW-PF objective function can provide a good sum SE aligned with acceptable performance for the worst UE.

\begin{figure*}[t]
    \centering
    \captionsetup{justification=centering}
    \begin{subfigure}{0.77\columnwidth}
        \centering
        \includegraphics[width=\columnwidth]{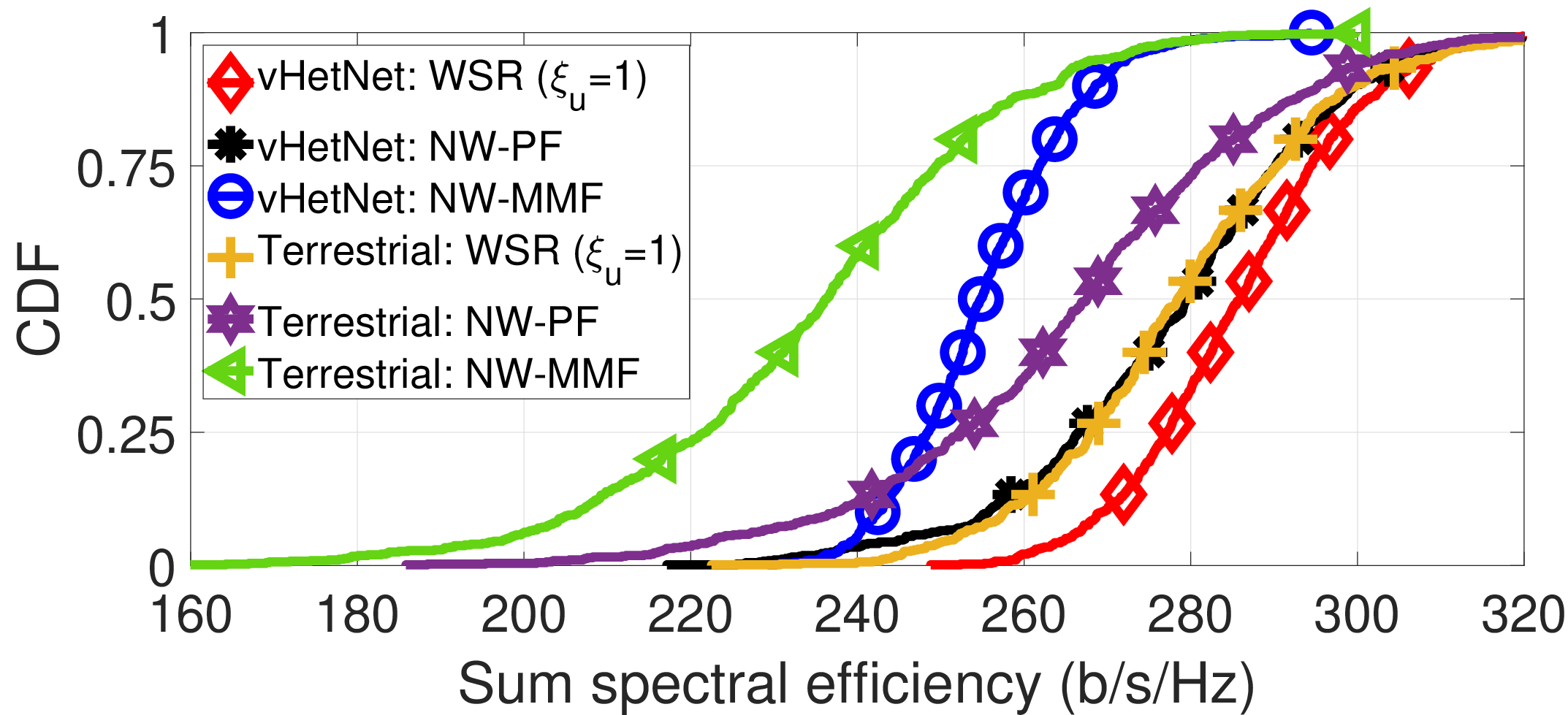}
        \caption{\small CDF of sum SE.}
        \label{fig_3:a}
    \end{subfigure}
    \begin{subfigure}{0.77\columnwidth}
        \centering
        \includegraphics[width=\columnwidth]{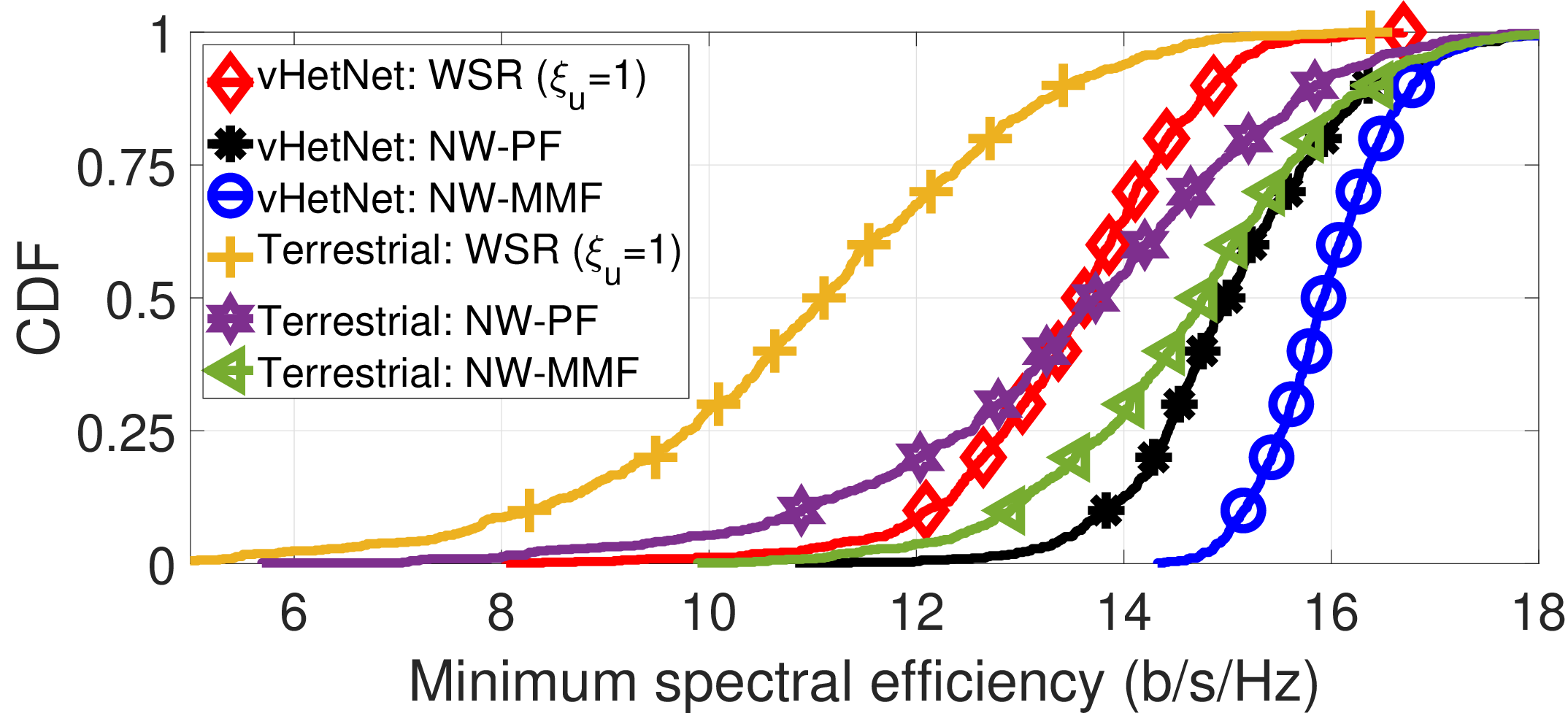}
        \caption{\small CDF of minimum SE.}
        \label{fig_3:b}
    \end{subfigure}
    \vspace{-1.5mm}
    \caption{\small Comparison between statistical behavior of SE in a vHetNet (4 MBSs + 1 HAPS) and standalone terrestrial network (5 MBSs).}
    \label{fig_3}
\end{figure*}

Fig.~\ref{fig_3} compares the performance of Algorithm~\ref{alg:JUBD}, using different objective functions, in vHetNet and standalone terrestrial network. Three observations can be obtained from Fig.~\ref{fig_3}. First, it can be observed that integrating HAPS with terrestrial networks improves the performance of the network. Second, it can be observed that the CDF curves in vHetNet are steeper than corresponding curves in the standalone terrestrial network, resulting in a more stable QoS in the vHetNet. Finally, as can be observed in Fig.~\ref{fig_3:a} and validated by Fig.~\ref{fig_3:b}, vHetNet provides higher SE values for the worst UEs. Particularly, an improved fifth percentile value can be observed from vHetNet in all algorithms. This observation can provide us with the conclusion that employing HAPS, to be integrated with the standalone terrestrial networks, can be considered a promising network architecture to improve the worst UE performance, and in particular the UEs that cannot establish a good channel condition in the terrestrial network, and therefore, receiving a low QoS.

\section{Conclusion}\label{Sec:Conclusion}
In this work, we formulated the joint user association and beamforming weights design (JUBD) problem for three objective functions (WSR, NW-PF, and NW-MMF). Since the formulated problems were non-convex, we employed reformulations and approximations to convert the problem to equivalent tractable SCA-based forms. Simulation results provided a comprehensive comparison between the statistical behavior of SE in different scenarios. The SE performance results under all three objective functions indicate that integrating HAPS with terrestrial networks improves the performance of the network. In addition, using NW-PF as the objective problem provides a good trade-off between achieved sum SE and minimum SE.

\vfill

\end{document}